\documentclass{article}

\usepackage{appendix}
\usepackage{amssymb}
\usepackage{amsmath}
\usepackage{chngcntr}
\usepackage[T1]{fontenc}
\usepackage{graphicx}
\usepackage[utf8]{inputenc}
\usepackage{natbib}
\usepackage{subcaption}
\usepackage{xcolor}

\title{Using child-woman ratios to infer demographic rates in historical populations with limited data}
\author{John Bryant \and Tahu Kukutai}
\date{}

\begin{document}
\maketitle

\begin{abstract}
  Data on historical populations often extends no further than numbers of people by broad age-sex group, with nothing on numbers of births or deaths. Demographers studying these populations have experimented with methods that use the data on numbers of people to infer birth and death rates. These methods have, however, received little attention since they were first developed in the 1960s. We revisit the problem of inferring demographic rates from population structure, spelling out the assumptions needed, and specialising the methods to the case where only child-woman ratios are available. We apply the methods to the case of M\={a}ori populations in nineteenth-century Aotearoa New Zealand. We find that, in this particular case, the methods reveal as much about the nature of the data as they do about historical demographic conditions.
\end{abstract}

\section{Introduction}
\label{introduction}

In the era of big data, it is easy to lose sight of the fact that, for most populations, for most of history, demographic data have been sparse at best. Even in recent decades, many countries only have scattered survey and administrative data, with a very occasional census \citep{un2022data}. Data on some populations, such as Indigenous populations, are scarcer still.

In sparse data settings, the data that do exist often take the form of counts of people by sex and age. Such data are among the least difficult to collect, and meet the most urgent needs of government administrators. In twentieth century censuses, the counts might be disaggregated by single-year or five-year age groups. In older data sources, the counts might only use broad age groups, such as counts of ``children'', ``men'', and ``women''.

During the 1960s,  Coale and Demeny, writing in a United Nations demographic methods manual \citep[][ch. 1]{united1967methods}, described how census data on population age structure could be used to infer fertility and mortality rates, especially if data on population growth were also available. The basic intuition behind these methods is that, since fertility and mortality rates determine age structure and growth rates, data on age structure and growth rates should allow some sort of inferences back to the underlying fertility and mortality rates. The simplest version of the Coale-Demeny methods makes the assumption that fertility and mortality rates have been constant, though Coale and Demeny show how the methods can be extended to deal with changing rates. The methods do not require data on births or deaths.

Since the 1960s, interest in these methods has waned among demographers studying contemporary populations, as more direct evidence on fertility and mortality levels, such as data from Demographic and Health Surveys, has become available. The IUSSP-UNFPA compendium of demographic methods, for instance, does not mention these methods \citep{moultrie2013tools}.

The data available to historical demographers has grown much more slowly since the 1960s. In historical demography, methods that can shed light on demographic rates without the need for direct data on fertility and mortality remain useful. Indeed, in Aotearoa New Zealand, data on age structure -- and particularly on the ratio between children and adult women -- have played a central role in the foundational research on M\={a}ori demographic history by Ian Pool \citep{pool1973estimates,pool2013te,pool2015colonization}.

In this paper, we revisit methods for inferring demographic rates from data on age structure and population growth. Following essentially the same logic as the early methods of Coale and Demeny, we show how the relationship between fertility, mortality, child-woman ratios, and population growth in stable populations can be used to make inferences about fertility and mortality rates. Whereas Coale and Demeny assume that census data on full age distribution can be obtained, however, we focus on the case where only counts of children versus adults are available, which allows us to develop a more direct, graphical approach than their original methods. We investigate whether these methods could shed light on demographic conditions experienced by M\={a}ori in nineteenth-century Aotearoa New Zealand. We find that the ability to draw out the implications of data on child-woman ratios and growth rates reveals as much about the limits of the data as it does about nineteenth century fertility and mortality.

\section{The relationship between fertility, mortality and child-woman ratios}

Throughout this paper, we define the child-woman ratio as the ratio of children aged 0--14 to women aged 15 and over. Alternative definitions, such as ones that restrict the denominator to women of reproductive age, are possible. Use of alternative definitions would not, however, change the main arguments of the paper.

The child-woman ratio can be used as an index of the age structure of a population. The higher the child-woman ratio, the wider the base of the population pyramid. The child-woman ratio increases when fertility increases, since higher fertility implies that more recent cohorts are larger than earlier ones. In high-mortality populations, the child-woman ratio also increases when mortality falls. The reason is that, in high-mortality populations, reductions in mortality entail larger absolute increases in the probability of survival for infants and young children than for other groups. Appendix~\ref{apsec:improve_life} presents an illustration of this phenomenon. A large absolute increase in the probability of survival for infants and young children has essentially the same effect on population structure as a large increase in fertility, in that it raises the number of `effective' births, or births of children who survive their first few years. The small size of the changes in survival probabilities in other age groups mean that the changes in these groups' probabilities have relatively little effect on overall population structure, so that the child-survival effect dominates.

To make these qualitative statements more precise, and to build a model of the relationship between fertility, mortality, and child-woman ratios, we need to make assumptions about the ways in which fertility and mortality rates vary. Following Coale and Demeny \citep{united1967methods}, we build on elementary results from stable population theory.

\begin{figure}
\centering
\includegraphics[width=0.8\textwidth]{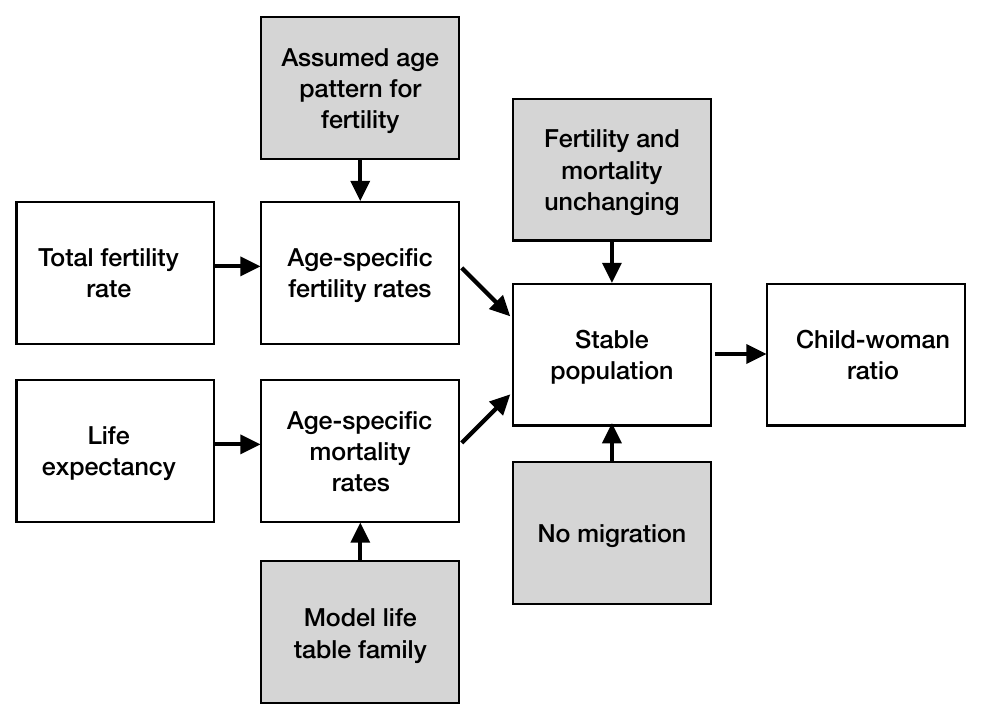} \caption{Deriving child-woman ratios from values for fertility and mortality. The inputs are the overall fertility and mortality rates, assumptions are shown in grey, the output is child-woman ratios, and all other boxes represent intermediate quantities.}
  \label{fig:diagram}
\end{figure}

Figure~\ref{fig:diagram} summarises our model. The inputs to the model are indicators for overall levels of fertility and mortality. Combining these indicators with assumptions about the age-sex-profiles of fertility and mortality yields age-sex-specific fertility and mortality rates. If we assume, in addition, that demographic rates have been constant for a long time, and that there is no migration, then the resulting population must be stable -- that is, the relative shares of each age-sex group are not changing, even if total population size is. Calculating the relative shares of each age-sex group is straightforward \citep{keyfitz2005applied}. Once we know the relative shares, we can calculate the child-woman ratio.

Our model uses the total fertility rate (TFR) as the summary indicator for overall fertility levels. The TFR is the average number of children that a woman would bear, under current fertility rates, if she survived to the end of the reproductive ages. For our initial calculations, we assume that age-specific fertility rates follow the `Booth standard' age pattern. The Booth standard underlies some widely-used methods for fertility estimation \citep{booth1984transforming,moultrie2013tools}. Although not shown in Figure~\ref{fig:diagram}, we also assume that there are 105 male births for every 100 female births.

Figure~\ref{fig:diagram} uses life expectancy as the summary indicator for overall mortality levels. In our initial calculations, we assume that age-specific mortality patterns follow the shape predicted by ``West'' model life tables \citep{coale1983models,jones2007demogr}. Our choice of a West model life table is based on the fact that it is a common default in studies of population dynamics, and that, according to \citet{pool2013te}, it is likely to be a reasonable approximation to nineteenth century M\={a}ori mortality patterns. In some of our analyses we work with life expectancies that are lower than those in published West model life tables. We derive life tables for these life expectancies that are consistent with West model life table age patterns by applying the Brass logit model to West model life tables \citep[][pp119-201]{preston2001demography}. (Specifically, we use the published life tables with the lowest life expectancies as our standard and vary the level parameter $\alpha$ until the implied life table has the target life expectancy.)

The realism of the assumption of unchanging demographic rates varies from population to population. It is, however, a convenient starting point, which we revisit later in the paper. The assumption of zero migration is, similarly, a convenient starting point.

\begin{figure}
  \begin{subfigure}{\textwidth}
\includegraphics[width=\textwidth]{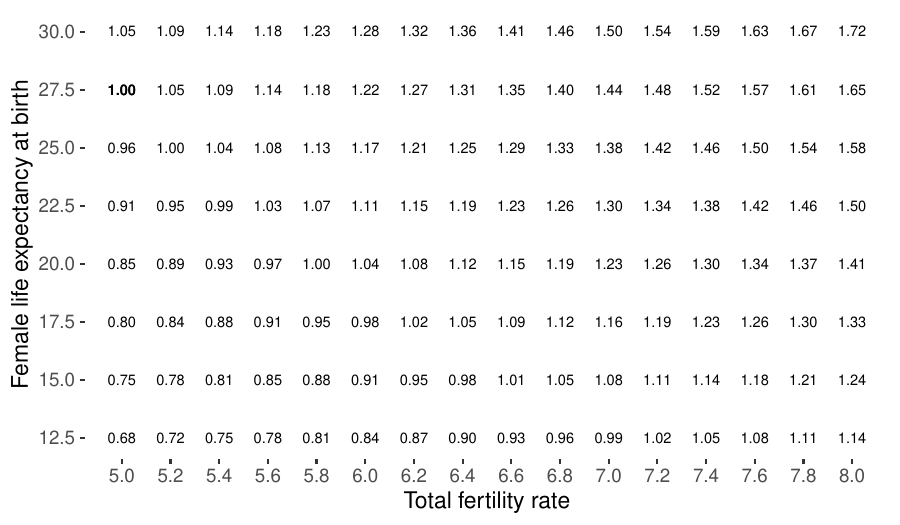} \caption{All child-woman ratios.}
\label{fig:fm2c-west-booth}
\end{subfigure}
\par\bigskip
\begin{subfigure}{\textwidth}
  \includegraphics[width=\textwidth]{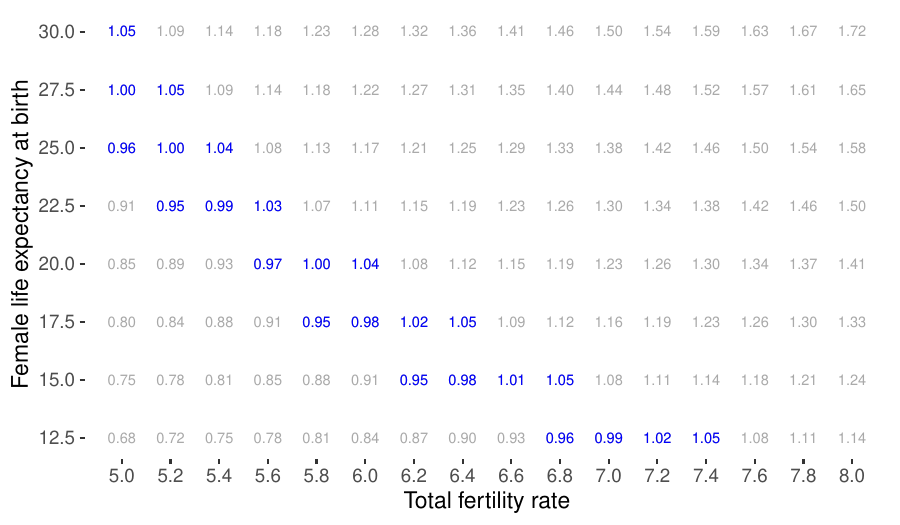} 
  \caption{Child-woman ratios in the range 0.95--1.05.}
  \label{fig:fm2c-west-booth-subset}
\end{subfigure}
  \caption{Child-woman ratios in stable populations, and associated values for fertility and mortality}
  \label{fig:fm2ct} 
\end{figure}

Figure~\ref{fig:fm2c-west-booth} displays the output from multiple applications of the model of Figure \ref{fig:diagram}. Each number in Figure \ref{fig:fm2c-west-booth} is a child-woman ratio obtained from a particular value for TFR, which varies along the horizontal axis, and for female life expectancy, which varies along the vertical axis. The number in bold towards the left of the figure shows, for instance, that a child-woman ratio of 1.00 is obtained from a TFR of 5.0 and a life expectancy of 27.5.

Comparing the values on right of Figure~\ref{fig:fm2c-west-booth} with the values on the left illustrates how higher fertility is associated with larger child-woman ratios. Comparing the values on the top of the figure with the values on the bottom illustrates how lower mortality (and hence higher life expectancy) is also associated with larger child-woman ratios.

\section{Inferring fertility and mortality from child-woman ratios}
  \label{sec:inferring}

When the assumptions behind Figure \ref{fig:fm2c-west-booth} are met, we can use the results from the figure to infer fertility and mortality rates from child-woman ratios. Given an observed value for child-woman ratio, we identify all combinations of TFR and life expectancy that could have given rise to this ratio. If, for instance, we observe a child-woman ratio of 1, we search through Figure~\ref{fig:fm2c-west-booth} to find all the 1s (including the case noted above) and record the associated values for TFR and life expectancy.

In practice, it is prudent to work with ranges of child-woman ratios rather than single values, to accommodate the fact that Figure~\ref{fig:fm2c-west-booth} only shows ratios at selected points, and as a way of acknowledging that the measurements of child-woman ratios are imprecise, and that the assumptions behind the model are unlikely to be fully met.

Figure~\ref{fig:fm2c-west-booth-subset} shows values of TFR and life expectancy that are associated with a child-woman ratios between 0.95 and 1.05. It shows, for instance, that a child-woman ratio of 0.95--1.05 can be produced by a TFR of 4--5 and a life expectancy of about 35, but also from a TFR of 6--7 and a life expectancy of about 15. It is impossible, using only information on child-woman ratios, to say which of
these possible combinations, or the many intermediate combinations, is more likely.

\section{Adding information on population growth rates}
  \label{sec:adding}

To narrow the range of possible values for fertility and mortality, we need to bring in more information besides child-woman ratios. One possibility is to bring in information on population growth rates. Estimates of population growth rates can be calculated from the same census data that are used to calculate child-woman ratios, provided that there are at least two censuses. Every stable population has an associated population growth rate, so the model of Figure~\ref{fig:diagram} can easily be extended to include the population growth rate as a second outcome alongside the child-woman ratio.

\begin{figure}
  \begin{subfigure}{\textwidth}
  \includegraphics[width=\textwidth]{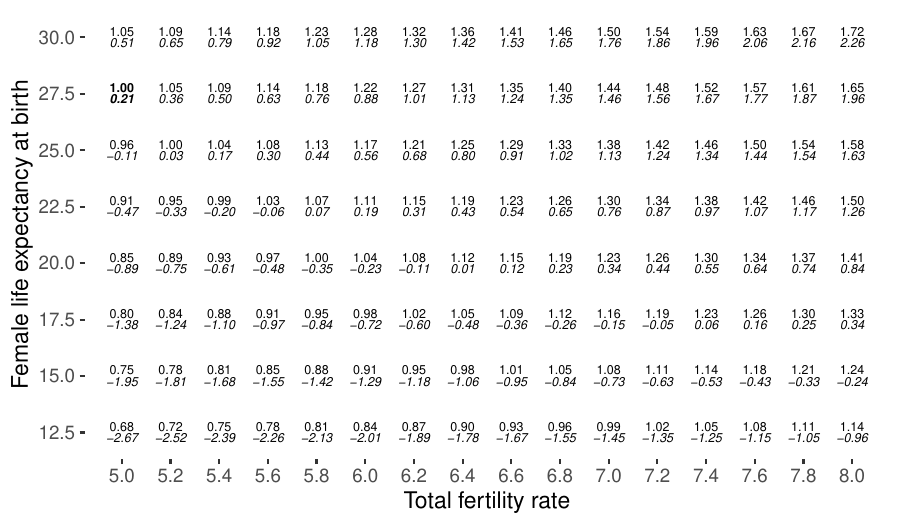} 
  \caption{All child-woman ratios and population growth rates.}
  \label{fig:fm2cg-west-booth-all}
\end{subfigure}
\par\bigskip
\begin{subfigure}{\textwidth}
  \includegraphics[width=\textwidth]{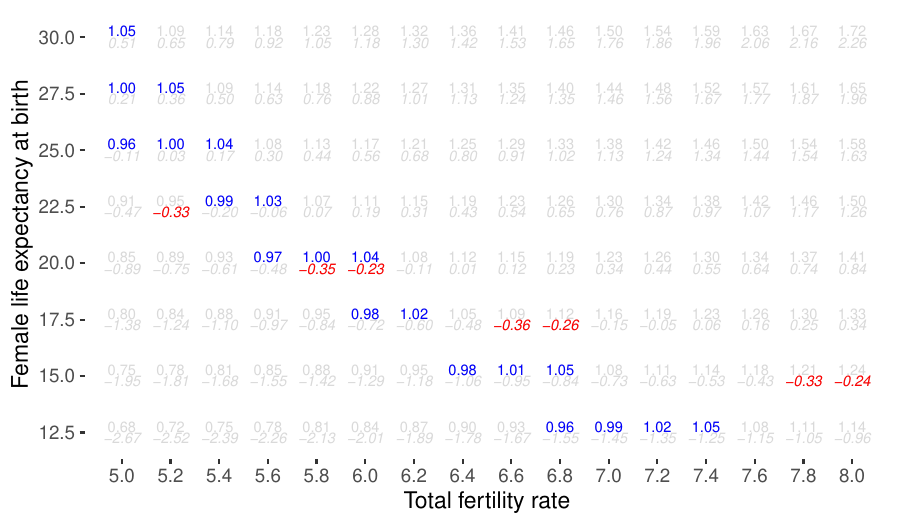}
  \caption{Child-woman ratios in the range 0.95--1.05 and population
    grow rates in the range -0.4\% to -0.2\%.}
  \label{fig:fm2cg-west-booth-subset}
\end{subfigure}
  \caption{Child-woman ratios and population growth rates in stable populations, and associated values for fertility and mortality. Population growth rates are shown in italics.}
  \label{fig:fm2cg-west-booth}
\end{figure}

Figure~\ref{fig:fm2cg-west-booth} shows results from applying the extended model. Figure~\ref{fig:fm2cg-west-booth} is identical to Figure~\ref{fig:fm2c-west-booth} except that, in addition to child-woman ratios, it also shows the annual population growth rates associated with each combination of TFR and life expectancy. The figure shows, for instance, that a TFR of 5.0 and a life expectancy of 27.5 leads
to a child-woman ratio of 1.0 and an annual population growth rate of 0.21\%.

The benefits from supplementing information on child-woman ratios with information on population growth rates are illustrated in Figure~\ref{fig:fm2cg-west-booth-subset}. Combinations of TFR and life expectancy that are consistent with child-woman ratios between 0.95 and 1.05 are shown in blue. Combinations that are consistent with with annual population growth rates between -0.4\% and -0.2\% are shown in red. Only two pairs of values fulfil both criteria: a TFR of 5.8 combined with a life expectancy of 20, and a TFR of 6.0 combined with a life expectancy of 20. We can conclude that in the hypothetical population in question, TFR is around 6 and life expectancy around 20.

\section{Sensitivity to model assumptions}
  \label{sec:sensitivity}

The assumptions behind our model of the relationship between fertility, mortality, child-woman ratio, and population growth rate will never be exactly met in any real population. In this section, we examine the effect of violations of the assumptions.

\subsection{Alternative age patterns for fertility and mortality}
  \label{sec:alternative}

We begin by investigating sensitivity to assumptions about age patterns of fertility and mortality. We examine how the relationship between fertility, mortality, and child-woman ratios, and between fertility, mortality, and population growth, changes when we depart from our original assumptions that age-specific fertility rates follow the Booth standard and that age-specific mortality rates follow the West model life table. Our alternative age pattern for fertility is the age pattern experienced by M\={a}ori in 1962 \citep{statsnz2020infoshare}. Our alternative age pattern for mortality is the South model life table \citep{coale1983models,jones2007demogr}. The alternative age patterns are graphed in Figure~\ref{fig:age_fm} in Appendix~\ref{apsec:alternative}. The 1962 M\={a}ori pattern has a sharper peak than the Booth standard, and South model life table has higher rates at the lowest and highest ages than the West model life table.

\begin{figure}
  \centering
  \includegraphics[width=0.8\textwidth]{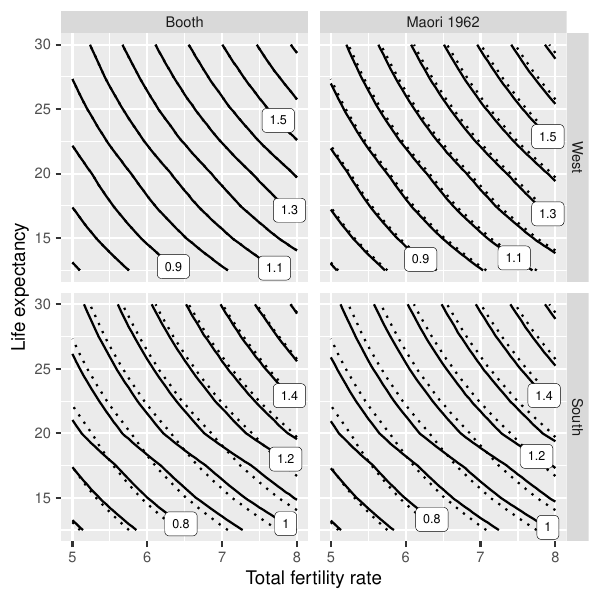} 
  \caption{Child-woman ratios under alternative assumptions about age patterns of fertility and mortality.  Each panel shows contour lines for child-woman ratios, that is, combinations of TFR and life expectancy that yield the same value for the child-woman ratio. The dashed lines represent the baseline combination of the Booth standard and West model life table.}
  \label{fig:sensitivity_cw}
\end{figure}

Figure \ref{fig:sensitivity_cw} shows how child-woman ratios vary with different combinations of age-patterns for fertility and mortality. Rather than showing numeric values for child-woman ratios, the figure shows contours lines. Each contour line in Figure \ref{fig:sensitivity_cw} depicts the combinations of TFR and life expectancy that are associated with a given child-woman ratio. To help with comparisons, the dashed lines in Figure~\ref{fig:sensitivity_cw} show child-woman ratios under the baseline combination of the Booth standard and West model life table.

Judging by the results in Figure \ref{fig:sensitivity_cw}, the relationship between TFR, life expectancy, and child-woman ratios is not at all sensitive to changes in the age pattern for fertility. The relationship is, however, sensitive to changes in the age pattern for mortality. For TFRs of 4.5--5.5, for instance, life expectancy needs to be 1 or 2 years higher under a South model life table to produce the same child-woman ratio that it does under a West model life table.

As discussed in Appendix~\ref{apsec:alternative}, similar conclusions apply to the relationship between TFR, life expectancy, and population growth, with changes in the age pattern for fertility rates having a much smaller impact on population growth rates than changes in the age pattern for mortality.

\subsection{Fertility and mortality rates changing over time}
  \label{sec:changing_fertility}

Our baseline model assumes that fertility and mortality rates have been fixed for a long period, and that the age-sex structure of the population has converged to its stable distribution. What happens if this assumption is violated? When fertility and mortality change, how quickly do child-woman ratios and population growth rates respond?

To assess the relationship between fertility and mortality rates, child-woman ratios, and population growth rates, we conduct simulation experiments. We look first at changes in fertility, examining two scenarios. In the first scenario, at time 0 the TFR increases instantaneously from 5 births to 5.25 births, and then remains at 5.25 births forever. In the second scenario, the TFR again starts at 5 births, but instead of just one increment of 0.25 births, there are five increments of 0.25 births, each five years apart.

\begin{figure}
  \centering
  \includegraphics[width=0.8\textwidth]{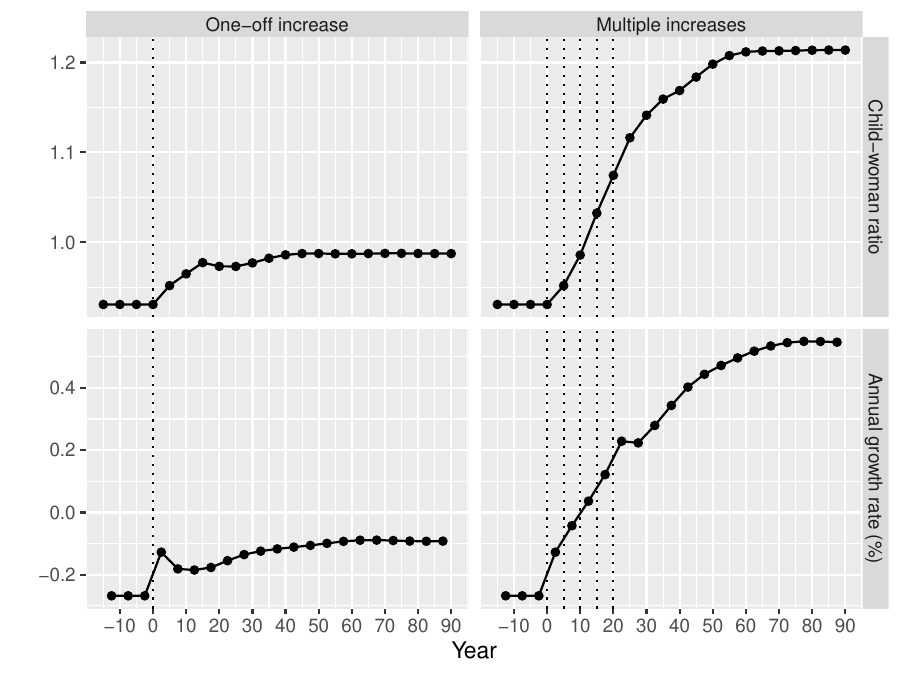}
    \caption{Changes in child-woman ratios and population growth in response to changes in fertility.}
    \label{fig:nonstable-fert}
\end{figure}

Results from these two scenarios are shown in Figure \ref{fig:nonstable-fert}. The first scenario is depicted in the two left-hand panels. In this scenario, child-woman ratios and population growth rates both take around 50 years to attain their new long-run values. The child-woman ratio gets most of the way towards its long-run value within the first 15 years. The population growth rate follows a more complicated path, with a rapid rise and fall followed by a slower rise.

The second scenario is depicted in the two right-hand panels. In this scenario, the adjustment periods for each increment in fertility overlap, producing a somewhat smoother transition. Once again, some of the adjustment happens fairly quickly, but it takes several decades for all the changes to work through.

Results from similar experiments with the effects of mortality changes are shown in Appendix~\ref{apsec:nonstable-mort}. The results are similar to those for fertility, though population growth adapts more quickly to a change in mortality than it does to a change in fertility.

The results from the simulation experiments suggest that when fertility and mortality rates are changing, the child-woman ratios and population growth rates can no longer be interpreted as providing information on demographic conditions at the time of the censuses. Instead, they should be interpreted as a lagged indicator of average conditions over the decades preceding the censuses. Changes in fertility and mortality rates have long-lasting effects on child-woman ratios and population growth rates, so current child-woman ratios and population growth rates as combined measures of fertility and mortality rates over the past generation or two.

\subsection{Migration}
  \label{sec:migration}

Migration into or out of an area might not be expected to affect child-woman ratios in that area, provided that mothers and their children migrate together. For migration to have absolutely no effect,
however, various other conditions need to be met: for instance, the probability that a woman migrates needs to be independent of the number of children she has given birth to.

In addition, unless inflows perfectly match outflows, migration will almost certainly affect population growth rates, even if mothers and children migrate together. Substantial migration can therefore be expected to push populations away from the combinations of indicators suggested by models based on fertility and mortality only.

\subsection{Measurement errors}
  \label{sec:measurement}

  The effect of under-coverage or over-coverage on the accuracy child-woman ratios and population growth rates is potentially complex. For instance, coverage errors only bias child-woman ratios to the extent that they vary between children and women, and only bias growth rates to the extent that they vary over time. In general, however, we would expect child-woman ratios to be more robust than population growth rates to variation in coverage errors. A 10\% difference in coverage rates between children and women will bias estimates of the child-woman ratio by about 10\%, but a 10\% difference in coverage rates between censuses can easily be larger than the true change in population size.

\section{Case study: Fertility and mortality of M\={a}ori in nineteenth-century Aotearoa New Zealand}

Quantitative evidence on the demographic effects on M\={a}ori of colonisation during the nineteenth century is patchy and difficult to interpret \citep{chapple2017new, anderson2017using}. We do not attempt assess the range of evidence, but simply to illustrate the contribution that analysis of on child-woman ratios and population growth can make.

Our input data are estimates of child-woman ratios and population growth from censuses in the years 1874, 1878, 1881, 1886, 1891, 1896, and 1901. We use data for all Aotearoa New Zealand. Rather than raw census counts, we use counts assembled by \citet[][Tables 5.2, 5.3]{pool2013te} that adjust for under-coverage, as estimated by Pool and other historical investigators.

The child-woman ratios and population growth rates have different time references, with child-woman ratios referring to census years, while population growth rates refer to inter-censal periods. To apply our methods, we need to bring the time references into alignment. We do this by aligning everything to the mid-point of each period, with the child-woman ratio equaling the average of the ratios at the two end points, and the population growth rate equaling the average growth rate for that period.  

\begin{figure}
  \includegraphics[width=\textwidth]{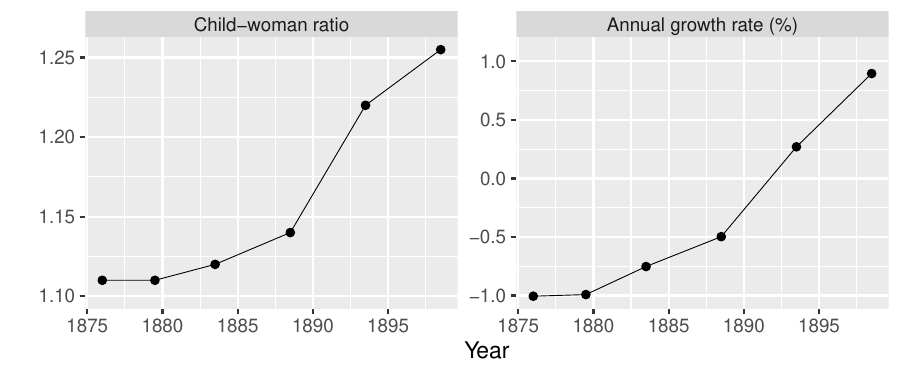}
  \caption{Estimates of child-woman ratios and annual population growth rates for M\={a}ori. Constructed from estimates presented in Table 5.2 and Table 5.3 of \citet{pool2013te}.}
    \label{fig:estimates}
\end{figure}

The resulting values are shown in Figure~\ref{fig:estimates}. As can be seen in the figure, over the 1870s and the 1890s the child-woman ratio increased from about 1.1 to about 1.25. Between the 1870s, the population growth rate  appears to have increased from about -1\% per year to about +1\% per year.

\begin{figure}
  \begin{subfigure}{\textwidth}
    \includegraphics[width=\textwidth]{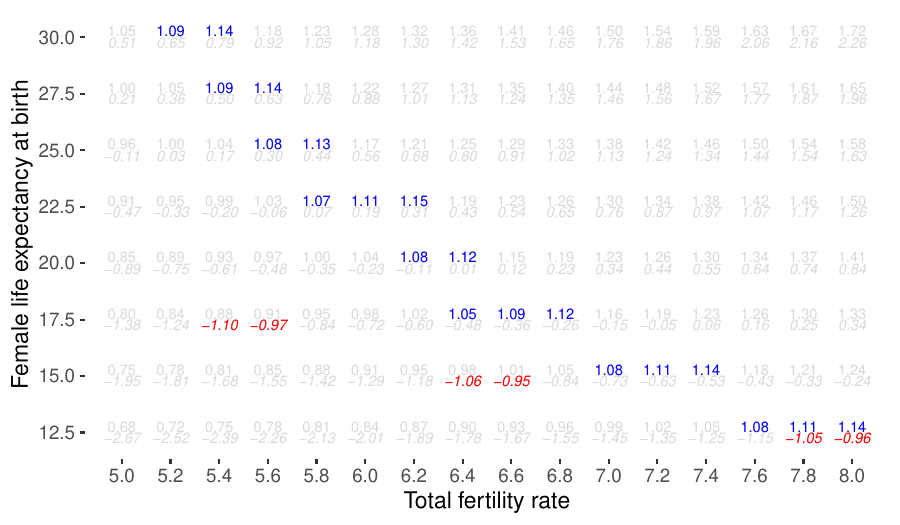}
    \caption{1874--1878}
  \end{subfigure}
  \begin{subfigure}{\textwidth}
    \includegraphics[width=\textwidth]{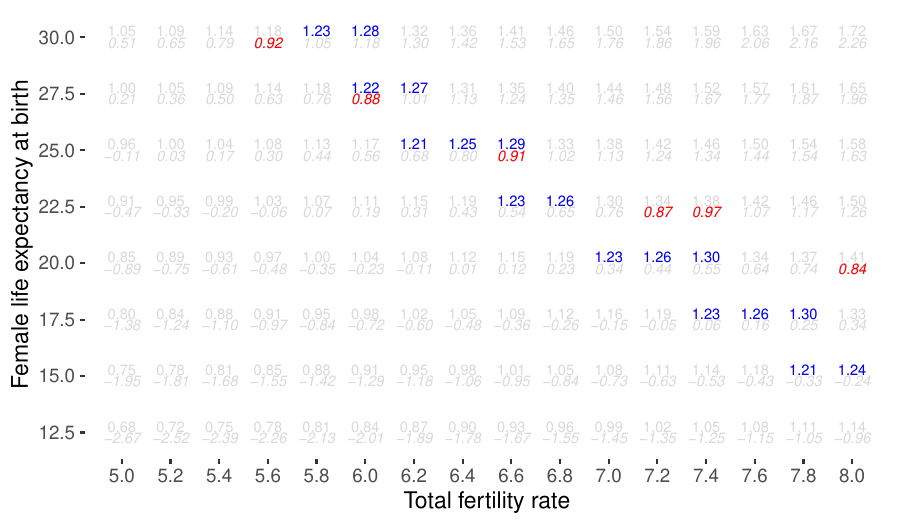}
    \caption{1896--1901}
  \end{subfigure}
  \caption{Life expectancy and TFRs implied by child-woman ratios and annual growth for M\={a}ori. The black numbers in the bottom right are derived with from the 1874 and 1878 censuses, and the black numbers near the top are derived from the 1896 and 1901 censuses.}
    \label{fig:map_fm2cg_nz}
  \end{figure}

Figure~\ref{fig:map_fm2cg_nz} shows the fertility and mortality rates implied by the estimated child-woman ratio and population growth rate for all M\={a}ori in Aotearoa New Zealand from the 1874--1878 censuses. TFR-life expectancy combinations consistent with the estimated child-woman ratio are shown in blue, and TFR-life expectancy combinations consistent with the estimated growth rate are shown in red. The two sets of combinations meet at the bottom right of the table, with a TFR of 7--8 and a life expectancy of 10--15.

The lower panel of Figure~\ref{fig:map_fm2cg_nz} shows the equivalent results derived from the 1896--1901 censuses. In this case, the two sets of combinations meet at the upper-middle part of the table, with a TFR of 6-7 and a life expectancy in the mid-to-high-20s.

  Although the later combination of TFR and life expectancy is plausible, the earlier combination is not. \citet[][p. 182]{rallu1992decline}, for instance, obtains life expectancies of 17--19 for the Marquesas in the early 20th century from good-quality birth and death registration data, but notes that these values were lower than they had been in the Marquesas during the late 19th century, and lower than those in India around 1900 during a plague epidemic.  It is unlikely that M\={a}ori mortality rates in 19th century Aotearoa New Zealand were significantly worse than those in the Marquesas. Moreover, given that conception and child birth both require some minimum level of health and security, it is extremely unlikely that very low life expectancies would have been paired with a TFR as high as 7--8.

It appears, then that the child-woman ratio from the earlier period, or the population growth rate, or both are wrong. Higher values for child-woman ratio and population growth rate would yield more plausible values for fertility and mortality. Finding some more plausible combination of child-woman ratio, population growth rate, TFR and life expectancy would, however, require additional information.

  The later combination of TFR and life expectancy is much more plausible. However, given the problems with the child-woman ratio and population growth rate in the earlier census data, we need to treat the child-woman ratio and population growth rate in the later census data with some suspicion. The estimates of a TFR of 6-7 and life expectancy in the mid-to-high-20s need to be treated as tentative.

\section{Discussion}

The original Coale-Demeny methods for analysing data on age-sex structure and population growth rates, and the more specialised methods for child-woman ratios that we have developed in this paper, are based on the relationship between fertility and mortality and population age-sex structure and growth rates. When data on age-sex structure and growth rates are sufficiently good, and when assumptions about matters such as age-patterns of mortality are approximately met, the data can be used to learn about fertility and mortality. However, as our case study shows, inference can also work in the opposition direction. If turns out that the population data implies fertility and mortality rates that are  implausible, then doubt is cast on the data itself.

The population data that we use in our case study have played a central role in analyses of the demographic impacts of colonisation in Aotearoa New Zealand \citep{chapple2017new, pool2015colonization}. Evidence that the population data, particularly for early periods, may have substantial errors is therefore important. Much work is still needed to reconstitute the population history of 19th century Aotearoa New Zealand.

One possible avenue for investigation is to distinguish more carefully between different areas, some of which are likely to have better data than others. Much of the available historical data for Aotearoa New Zealand can be disaggregated geographically. Enumerators' reports provide at least some information on how reliability varies across areas and across periods \citep{statsnz2022census}. A combination of demographic analysis, involving measures such as child-woman ratios and growth rates, with investigation of historical sources might allow researchers to distinguish between areas where demgographic reconstructions were feasible and areas where it was not. This would yield new, more detailed information on the demographic impacts of colonisation, and give a clearer picture of the strengths and weaknesses of the available evidence.

\clearpage

\bibliography{references}

\newpage

\begin{appendices}
\counterwithin{figure}{section}

\section{How improvements in life expectancy affect survival probabilities of different age groups}
  \label{apsec:improve_life}

  \begin{figure}[!hb]
    \centering
  \includegraphics[width=0.8\textwidth]{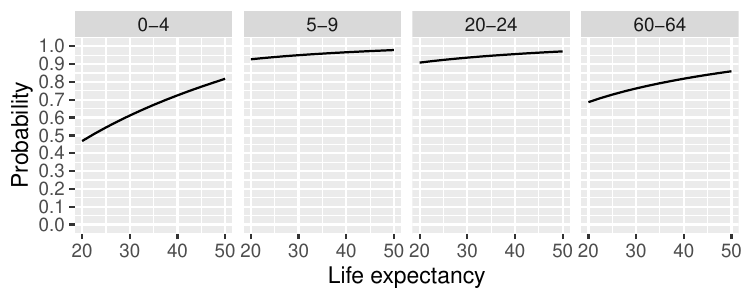} \caption{How survival probabilities at selected ages vary with life expectancy, under West model life tables for females.}
  \label{fig:survival}
\end{figure}

Figure~\ref{fig:survival} shows how the probability of surviving from the start to the end of the age groups 0--4, 5--9, 20--24, and 60--64 change as the level of overall mortality falls from very high to moderate. The probabilities come from West model life tables for females \citep{coale1983models, jones2007demogr}, which capture experiences across a wide range of countries. The absolute increases in survival probabilities for 0--4 year olds are much larger than the absolute increases for 5--9 year olds and 20--24 year olds. The absolute increases in survival probabilities for 60--64 year olds are somewhat larger, but older people constitute a small proportion of the population in high-mortality populations, so changes in the size of this group has only a small effect on child-woman ratios.

\clearpage

\section{Alternative assumptions about age patterns of fertility and mortality}
\label{apsec:alternative}

\begin{figure}[!hb]
  \includegraphics[width=\textwidth]{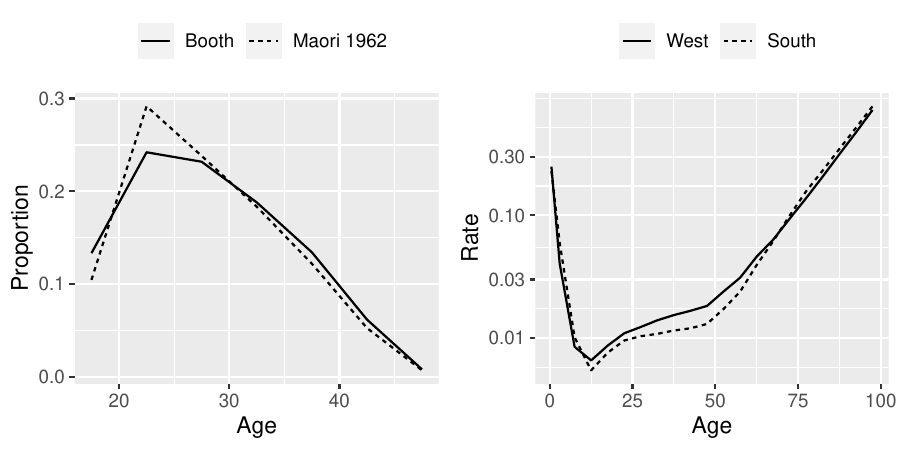} \caption{Original and alternative age patterns for fertility and mortality. The left panel compares our original age-pattern for fertility, based on the Booth standard, with an assumption based M\={a}ori fertility rates in 1962. The right panel compares our original age-pattern for mortality, based on the West model life table, with the age-pattern from the South model life table.}
  \label{fig:age_fm}
\end{figure}

Figure~\ref{fig:age_fm} shows our baseline assumptions about age patterns for fertility and mortality, versus the alternative assumptions that we use for sensitivity testing in Section~\ref{sec:sensitivity}. Compared with the Booth standard, the M\={a}ori 1962 has a sharper peak at young ages. Compared with the West age pattern, the South pattern has higher childhood and old-age mortality, and lower middle-age mortality.

\begin{figure}
  \centering
  \includegraphics[width=0.8\textwidth]{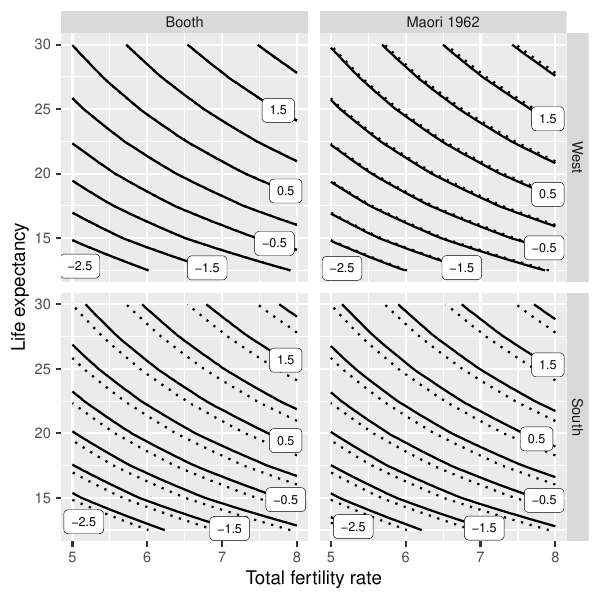} 
  \caption{Annual population growth rates (percent) under alternative assumptions about age patterns of fertility and mortality.  Each panel shows contour lines for population growth rates, that is, combinations of TFR and life expectancy that yield the same value for the growth rate. The dashed lines represent the baseline combination of the Booth standard and West model life table.}
  \label{fig:sensitivity_gr}
\end{figure}

Figure~\ref{fig:sensitivity_gr} is identical to Figure~\ref{fig:sensitivity_cw} , except that it shows population growth rates rather than child-woman ratios. The results are essentially the same as those for child-woman ratios. Switching to a new assumption about the age-pattern of fertility has a negligible effect on the relationship between fertility, mortality, and the population growth rate, while switching to a new assumption about the age-pattern of mortality as a much more noticeable effect.

\clearpage

\section{Allowing for changes in mortality rates}
  \label{apsec:nonstable-mort}

\begin{figure}[!hb]
    \centering
  \includegraphics[width=0.8\textwidth]{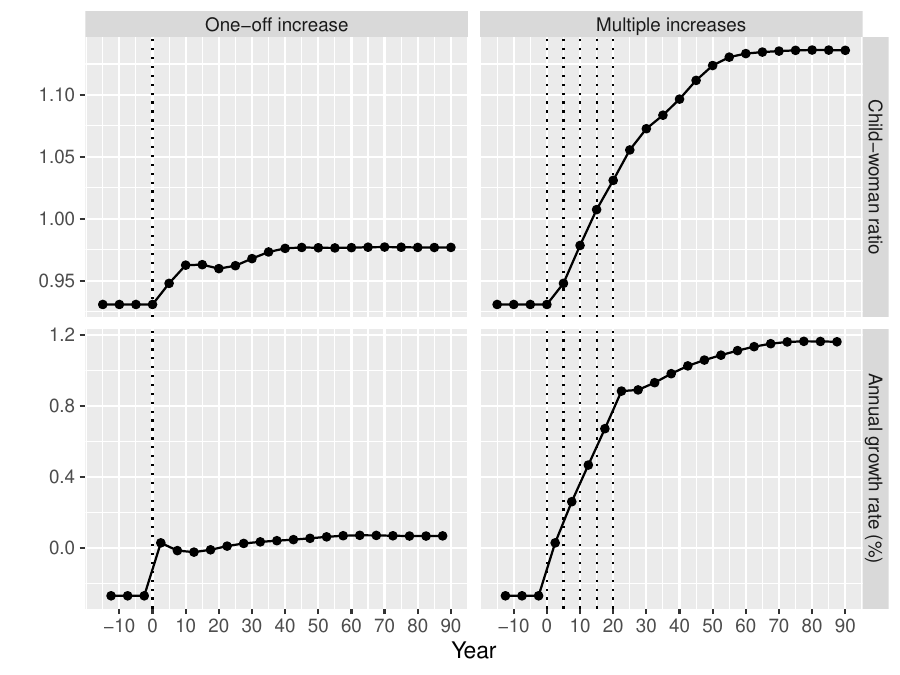}
  \caption{\emph{Changes in child-woman ratios and population growth in response to changes in mortality.}}
  \label{fig:nonstable-mort}
\end{figure}

Figure~\ref{fig:nonstable-mort} is identical to Figure~\ref{fig:nonstable-fert}, except that it uses life expectancy in place of the TFR. In the two left-hand panels of Figure~\ref{fig:nonstable-mort}, life expectancy starts at 30 and then increases to 32.5. In the two right-hand panels, life expectancy starts at 30 and then increases by 2.5 years ever 5 years. The results are similar to those for the TFR, except that the adjustment process is somewhat faster.

\end{appendices}

\end{document}